\documentclass{article}
\usepackage{amsmath}
\usepackage{amsfonts}
\usepackage{amssymb}
\usepackage{graphicx}
\usepackage[mathscr]{eucal}
\usepackage{hyperref}

\begin{document}

\title{Entropic Dynamics on Curved Spaces\thanks{Presented at MaxEnt 2015, the 35th International Workshop on Bayesian Inference and Maximum Entropy Methods in Science and Engineering
(July 19--24, 2015, Potsdam NY, USA). } }
\author{Shahid Nawaz\thanks{%
corresponding auther: snafridi@gmail.com}, Mohammad Abedi\thanks{mabedi@albany.edu} \ and Ariel Caticha\thanks{%
ariel@albany.edu} \\
{\small Department of Physics, University at Albany-SUNY, }\\
{\small Albany, NY 12222, USA.}}
\date{}
\maketitle
\begin{abstract}
Entropic dynamics is a framework in which quantum theory is derived as an application of entropic methods of inference. Entropic dynamics on flat spaces has been extensively studied. The objective of this paper is to extend the entropic dynamics of $N$ particles to curved spaces. The important new feature is that the displacement of a particle does not transform like a vector because fluctuations can be large enough to feel the effects of curvature. The final result is a modified Schr\"odinger equation in which the usual Laplacian is replaced by the Laplace-Beltrami operator.
\end{abstract}
\section{INTRODUCTION}
Entropic dynamics (ED) is a framework in which the laws of dynamics, such as quantum mechanics, are derived as an example of entropic inference. In previous publications ED has been used to formulate the non-relativistic quantum mechanics of particles moving in flat Euclidean space \cite{caticha2012}-\cite{Caticha2015}.
The objective of this paper is to extend the entropic dynamics of $N$ particles to curved spaces. This is an important preliminary step toward an entropic dynamics of gravity. Two other related contributions appear in \cite{Ipek.Caticha, catichaGIG}. 

The formalism developed here is a straightforward extension of previous work. The main problem to be tackled is technical: in a curved space the displacement of a particle does not transform like a vector. The reason is that in entropic dynamics, just as in other forms of dynamics that involve Brownian motion such as Nelson's stochastic mechanics \cite{Nelsonqf}, the fluctuations in the particle's motion tend to be so large that they can feel the curvature of the underlying space. For a single particle this technical problem was tackled in \cite{Shahidspin}; here we extend the formalism to several particles. In its early formulations ED involved a number of assumptions about auxiliary variables and about the form of the Hamiltonian and of the quantum potential that were not sufficiently justified. Later developments in \cite{Caticha2014} showed that the auxiliary variables were in fact unnecessary. Furthermore, in \cite{CBR2014} it was shown that the requirement that ED describe a non-dissipative diffusion leads naturally to a Hamiltonian dynamics, and that the tools of information geometry can be used to justify  the metric for the $N$-particle configuration space and the functional form of the quantum potential. This latter development is crucial to our current purposes: just specifying the geometry of the curved space in which a single particle moves is not sufficient to specify the geometry of the configuration space for $N$ particles. 
In this work all these improvements -- the elimination of auxiliary variables, the derivation of Hamiltonian dynamics, and the use of information geometry -- are implemented for the ED of $N$ particles moving in a curved space. The final result is a modified Schr\"odinger equation that takes into account the effects of curvature.

\section{ENTROPIC DYNAMICS} \label{SM-curved}

Just as in any framework for inference it is crucial that we start by identifying the variables that we wish to infer. For the present paper the subject of our inference is the positions of particles. Perhaps the central feature of ED -- that which establishes the subject matter -- is the assumption that although these positions are unknown they have definite values. This is in marked contrast with the standard Copenhagen interpretation in which observables do not have actual values until elicited by an act of measurement. 

We consider $N$ particles, each of which lives in a curved space $\mathscr{X}$ of $d$-dimensions and metric $h_{ab}$. The configuration space is $\mathscr{X}_N$=$\mathscr{X} \times \mathscr{X} \ldots \times \mathscr{X}$ and has dimension  $n=Nd$. 

In order to find the probability distribution for the particles' positions $\rho (x,t)$ we will first use the method of maximum entropy to determine the probability $P(x^{\prime} | x)$ that the particles take an infinitesimally short step from an initial position $x\in\mathscr{X}_N$ to a final position $x^\prime\in\mathscr{X}_N$. Having determined the transition probability for a short step we will later iterate this result to find $\rho (x,t)$. To find $P(x^{\prime} | x)$ we must first identify a prior  $Q(x^{\prime}| x)$ and the appropriate constraints and then we maximize the joint entropy,
\begin{equation}
\mathscr{S}\left[ P,Q\right] =-\int\! d^nx^{\prime}  P\left( x^{\prime }|x\right) \log \frac{P\left( x^{\prime
}|x\right) }{Q\left( x^{\prime }|x\right) }\, .
\label{RelEnt-curved}
\end{equation}

\paragraph{The Prior}
The prior probability expresses the uncertainty about which $x^{\prime}$ to expect before any information is taken into account. In the state of extreme ignorance we have no idea of where the particles will move and knowing the position of one will tell us nothing about the positions of the others. The prior is a product of the uniform priors for the individual particles. It is given by
\begin{equation}
Q(x^\prime | x) = \prod_{i=1}^{N} Q (x^\prime_i | x_i) \, ,
\end{equation}
where, for instance, $x_i$ is the initial position of $i^{th}$ particle. Furthermore, the individual priors are uniform which means that equal volumes are assigned equal probabilities,
\begin{equation}
Q(x^\prime_i | x_i) = h^{1/2} (x^\prime_i) ~,
\label{Prior}
\end{equation}
where $h = \det h_{ab}$ . (We need not worry about normalization because it has no effect on the final result.)\\

\paragraph{The Constraints}
The information about the motion is introduced through the constraints. The first constraint deals with continuity of motion. This means that the motion can be analyzed as a sequence of infinitesimally short steps from $x_i^a$ to  $x^{\prime a}_i=x^{a}_i+\Delta x^{a}_i$ ($i$ is the particle index and $a$ is its spatial coordinate index which runs from 1 to $m$). This information is incorporated by the following constraint  
\begin{equation}
\left\langle h_{ab} \Delta x^{a}_i \Delta x^{b}_{i}\right\rangle \ =\kappa_i\,,  \label{Constraint2-curved}
\end{equation}%
where $\kappa_i$ is a small constant that is eventually allowed to tend to zero. These $N$ constraints lead to a motion in which the particles diffuse isotropically and independently of each other. In order to introduce some directionality and also to account for entanglement effects we introduce one additional constraint that acts on the configuration space. We impose that the expected displacement in the direction of the gradient of a certain "potential" $\phi$ satisfies 
\begin{equation}
\sum_{i=1}^{N} \langle \Delta x^a_i \rangle \frac{\partial \phi}{\partial x^a_i}=\kappa^\prime\,,\label{potential-constraint}
\end{equation}
 where $\kappa^\prime$ is another small constant.

Finally maximize eq.~(\ref{RelEnt-curved}) subject to the  constraints eqs.~(\ref{Constraint2-curved}, \ref{potential-constraint}) and normalization. We obtain
\begin{equation}
P(x^{\prime}|x)=\frac{\prod_{i=1}^{N}  h^{1/2}(x^\prime_i)}{\zeta(x,\alpha_1, \alpha_2,\cdots,\alpha_N,\alpha^\prime)}\exp\ {\left[-\sum_{i=1}^{N} \left(\frac{1}{2} \alpha_i   h_{ab}(x_i) \Delta x^a_i\Delta x^b_{i}-\alpha^\prime \Delta x^a_i \frac{\partial \phi}{\partial x^a_i}\right)\right]}\,, \label{TransProb1-curved}
\end{equation}
where $\alpha_i$  ($i=1\cdots N$) and $\alpha^\prime$ are Lagrange multipliers and $\zeta$ is a normalization constant. There is a potential problem with eq. (\ref{TransProb1-curved}) because as we shall see below in eq. (\ref{tayler}) neither coordinate differences such as $\Delta x^a_i$ nor their expected values $<\Delta x^a_i>$ and $<\Delta x^a_i \, \Delta x^b_i>$ are covariant which means that the constraints eq. (\ref{Constraint2-curved}) and eq. (\ref{potential-constraint}) are not covariant. However, as we shall see later, we require the transition probability $P(x'|x)$ in the limit of short steps or $\alpha_i \rightarrow \infty$. It is only in this limit that we will later achieve manifestly covariant results.  

\paragraph{Entropic Time}
The concept of time is closely connected with motion and change \cite{Caticha2001}. In ED motion is described by the transition probability, eq.~(\ref{TransProb1-curved}), that describes infinitesimally short steps. Larger finite changes will be obtained as the accumulation of many short steps.
To construct time we must specify what we mean by an instant, how instants are ordered, and the interval or duration between them \cite{Catichatime2011}. In ED an instant is defined by the information required to generate the next instant. In our case this information is given by the probability density in configuration space $\mathcal{P}(x,t)$. Thus, in ED an instant $t$ is represented by a probability distribution. The new distribution $\mathcal{P}(x^\prime ,t^\prime)$ at the next instant $t^\prime$ can be constructed using the transition probability given by eq.(\ref{TransProb1-curved})
\begin{equation}
\mathcal{P} (x^{\prime},t^{\prime})=\mathop{\textstyle{\int}}\!d^nx\, P(x^{\prime}|x)  \mathcal{P} (x,t) \,. \label{rho1-curved}
\end{equation}
 
Next we introduce the notion of duration -- the interval $\Delta t$ between successive instants. For non-relativistic quantum mechanics we want to construct a time that is Newtonian in the sense that the time interval is independent of $x$ and $t$. This can be done through an appropriate choice of $\alpha_i$,
\begin{equation}
\alpha_i =\frac{m_i}{\eta \Delta t} \,, \label{alpha-curved}
\end{equation}
where $m_i$ is a particle dependent constant that will turn out to be the mass of the $i^{th}$ particle and $\eta$ is an overall constant that fixes the units of time relative to those of mass and length.

\paragraph{The Information Metric of Configuration Space}
The configuration space  $\mathscr{X}_N$  is a smooth topological manifold. Although we have introduced a metric for the single particle space $\mathscr{X}$ we have not thus far introduced a metric for  $\mathscr{X}_N$. The natural candidate is the information distance between two neighboring distributions $P(x^\prime | x)$ and $P(x^\prime | x+dx)$ which is given by the information metric
\begin{equation}
g_{AB} = C \int d^n x^{\prime} P(x^{\prime} | x) \frac{ \partial \log P(x^{\prime} | x)}{\partial x^A} \frac{ \partial \log P(x^{\prime} | x)}{\partial x^B} \,,\label{infometric}
\end{equation}
where $C$ is an undetermined scale factor and $\partial / {\partial x^A} =  \partial/{\partial x^a_i} $. The capitalized indices such as $A = (i,a)$ and $B = (j,b)$ denote both the particle index and its spatial coordinates. Substituting eq.(\ref{TransProb1-curved}) into eq. (\ref{infometric}) yields
\begin{equation}
g_{AB} = \frac{C m_i}{\eta \Delta t} \delta_{ij}\, h_{ab} \, .
\end{equation}

Since in what follows the metric $g_{AB}$ will always appear multiplied by the combination $C/{\eta \Delta t}$, it is convenient to introduce an effective metric
\begin{equation}
M_{AB}(x) = \frac{\eta \Delta t}{C} g_{AB}(x) = m_i  \delta_{ij}\, h_{ab}(x_i) \, \label{M} ,
\end{equation}
which we will call the mass tensor because that is what it is in the special case of flat spaces where $M_{AB}$ is independent of $x$. \\

\paragraph{The Transition Probability}
Substituting eq.(\ref{M})  into eq.(\ref{TransProb1-curved}) the transition probability becomes
\begin{equation}
P(x^{\prime}|x)=\frac{ M^{1/2}(x^{\prime})}{\zeta^\prime (x,\alpha_1,\alpha_2,\cdots,\alpha_N,\alpha^\prime)}\exp\ {\left[- \left(\frac{1}{2 \eta \Delta t}  M_{AB}(x) \Delta x^A\Delta x^B-\alpha^\prime \Delta x^A \frac{\partial \phi}{\partial x^A}\right)\right]}\, \label{trans},
\end{equation}
where $\zeta^\prime $ is the new normalization constant and $M(x^\prime)=\det {M_{AB}(x^\prime)}$ . The detailed analysis of $\alpha^\prime$ appears in \cite{D.B A.C MaxEnt15}. Here we we just note that for our current purposes $\alpha^\prime$ can be absorbed into $\phi$ which amounts to setting $\alpha^\prime = 1$ .\\

\paragraph{Analysis}
Next we want to express a generic displacement $\Delta x^A$ in terms of an expected drift $\left<\!\Delta x^A\! \right>$ plus a fluctuation $\Delta w^A$. The difficulty is that due to the determinant in the pre-exponential factor the distribution (\ref{trans}) is not Gaussian which makes it difficult to calculate expected values exactly. To solve this problem we note that as $\Delta t \rightarrow 0$, $P(x^\prime|x)$ probes a very localized region of $\mathscr{X}_N$. This suggests a Taylor expansion about $x$. Unfortunately since we deal with a Brownian motion to account for fluctuations we need to keep track of quadratic terms $\Delta x^A \Delta x^B$ which are affected by curvature. A Brownian particle does not follow a smooth trajectory. Its fluctuations will make it probe the curvature effects in a local neighborhood of $x$.

 The configuration space $\mathscr{X}_N$ is a curved space which is locally like $\mathbb{R}^n$. It is convenient to express the transition probability in normal Cartesian  coordinates at $x_p$. In normal coordinates (NC) the metric tensor in the vicinity of a point $x_p$ is approximately that of flat Euclidean space, i.e.
\begin{equation} 
h_{ab} = \delta_{ab} , \hspace{5mm} \left. \frac{\partial h_{ab}}{\partial x^a} \right|_{x_p} = 0 \, .
\end{equation}
Therefore for the configuration space metric we have
\begin{equation}
M_{A^\prime B^\prime}(x_p) =\gamma_{A^\prime B^\prime},\quad \textrm{with}\quad \gamma_{A^\prime B^\prime}=m_i \, \delta_{ij}\, \delta_{ab} \label{gamma} \,,
\end{equation}
we use primed indices for NC and unprimed for generic coordinates. The first derivative vanishes
\begin{equation}
 \left.\frac{\partial M_{A^\prime B^\prime}}{\partial x^{C^\prime}} \right|_{x_p} = 0 \, \hspace{5mm} \textrm{but} \hspace{3mm}  \left.\frac{\partial^2 M_{A^\prime B^\prime}}{\partial x^{C^\prime} \partial x^{D^\prime}} \right|_{x_p} \neq 0 \, .
\end{equation}
In NC $\det{M}= \det{\gamma}=\prod_{i=1}^{N} (m_i)^d$ which can be absorbed into a new normalization constant. Notice that the second order term in the Taylor expansion of $M_{A^\prime B^\prime}$ is proportional to $\Delta x^{A^\prime} \, \Delta x^{B^\prime}$ which is of order $\Delta t$ and can therefore be neglected relative to terms of order $\Delta t^0$ and $\Delta t^{1/2}$ in the exponent. The transition probability becomes Gaussian
\begin{equation}
P_{NC}\left( x^{\prime }|x\right) = \frac{1}{Z}\exp\ \left[- \frac{1}{2 \eta \Delta t}  \gamma_{A^\prime B^\prime}(x) (\Delta x^{A^\prime} -  \langle \Delta{x}^{A^\prime} \rangle ) \, ( \Delta x^{B^\prime} - \langle \Delta {x}^{B^\prime} \rangle) \right] \,,
\end{equation}
where $Z$ is the normalization constant in normal coordinates.
The displacement $\Delta x^{A^\prime}$ is expressed as the expected drift  plus a fluctuation
\begin{equation}
\Delta x^{A^\prime} = \langle \Delta {x}^{A^\prime} \rangle+\Delta w^{A^\prime} ~ \label{Displacement1-NC},
\end{equation}
where 
\begin{equation}
\langle\Delta {x}^{A^\prime} \rangle=\eta \, \Delta t \,  \gamma^{A^\prime B^\prime} \frac{\partial \phi}{\partial x^{B^\prime}} ~, \label{mean-NC}
\end{equation}
and
\begin{equation}
\langle\Delta w^{A^\prime}\rangle =0\,, \quad \textrm{and} \quad \langle\Delta w^{A^\prime} \Delta w^{B^\prime}\rangle = {\eta \, \Delta t} \, \gamma^{A^\prime B^\prime}\,. \label{fluctuations-NC}
\end{equation}
Having calculated the diffusion process in the special NC coordinates we now transform back to generic coordinates. We Taylor expand $ x^{A^\prime}(x^A)$ about point $x_p$. ( See e.g.\ \cite{Nelsonqf, Shahidspin} ).

\begin{equation}
\Delta x^{A^\prime} = \frac{\partial x^{A^\prime}}{\partial x^A} \Delta x^{A} + \frac{1}{2} \Delta x^{A} \Delta x^{B} \frac{\partial^2 x^{A^\prime}}{\partial x^{A} \partial x^{B}} + \cdots \, , \label{tayler}
\end{equation}
The displacement $\Delta x^{A^\prime}$ involves fluctuations of the order of $\mathcal{O}({\Delta t}^{1/2})$, therefore the second order term must be included. This shows that $\Delta x^{A^\prime}$ does not transform as a vector. To calculate the correction term it is convenient to introduce a vector $\tilde{\Delta} x^{A}$ that coincides with $\Delta x^{A^\prime}$ in NC and in generic coordinates it is given by
\begin{equation}
\tilde{\Delta} x^{A} = \frac{\partial x^{A}}{\partial x^{A^\prime}} \Delta x^{A^\prime} \, . \label{tildex}
\end{equation}
 Substituting eq. (\ref{tayler}) into eq. (\ref{tildex}) we get
\begin{equation}
\tilde{\Delta}x^{A} = \Delta x^{A} + \frac{1}{2} \Delta x^{B} \Delta x^{C} \left( \frac{\partial^2 x^{A^\prime}}{\partial x^{B} \partial x^{C}} \frac{\partial x^{A}}{\partial x^{A^\prime}} \right) \, .
\end{equation}
The second term is related to the transformation of Christoffel's symbols 
\begin{equation}
\Gamma^{A}_{B C} = \frac{\partial x^{A}}{\partial x^{A^\prime}} \frac{\partial x^{B^\prime}}{\partial x^{B}} \frac{\partial x^{C^\prime}}{\partial x^{C}} \Gamma^{A^\prime}_{B^\prime C^\prime} + \frac{\partial x^{A}}{\partial x^{A^\prime}} \frac{\partial^2 x^{A^\prime}}{\partial x^{B} \partial x^{C}} \, .
\end{equation}
Since $\Gamma^{A^\prime}_{B^\prime C^\prime} = 0 $ in NC, then
\begin{equation}
\Delta x^A = \tilde{\Delta} x^A - \frac{1}{2} \Gamma^A_{BC} \Delta x^B \Delta x^C \, \label{delx} .
\end{equation}
We can rewrite this equation as
\begin{equation}
\Delta x^{A}=b^{A}(x)\Delta t+\Delta w^{A}\,,  \label{Displacement2-curved}
\end{equation}
where $b^A$ is the drift velocity
\begin{equation}
b^{A}(x)=\tilde{b}^{A}(x)-\frac{\eta}{2}M^{BC} \Gamma^A_{BC} ~,\quad\textrm{with}\quad \tilde{b}^{A}(x)= {\eta}M^{AB}\partial_{B}\phi\, , \label{mean2-curved}
\end{equation}
where $M^{AB}$ is the inverse of $M_{AB}$. Notice that $b^A$ does not transform like a vector but $\tilde{b}^A$ does. And the fluctuation

\begin{equation}
\langle\Delta w^{A}\rangle =0\, \ \ \textrm{and} \quad \langle\Delta w^{A}\Delta w^{B}\rangle = \eta M^{AB}\Delta t \,. \label{fluctuations2-curved}
\end{equation} 
To conclude this section we rewrite the transition probability eq.  (\ref{trans}) in generic coordinates in the limit $\Delta t \to 0$,
\begin{equation}
P(x^\prime|x) = \sqrt{M(x^\prime)}\  \tilde{P}(x^\prime|x) = \frac{\sqrt{M(x^\prime)}}{\tilde{\zeta}} \exp \left[-\frac{1}{2 \eta \Delta t} \, M_{AB} \, (\tilde{\Delta} x^A - \eta \, \Delta t \, \partial^A \phi) \, (\tilde{\Delta} x^B - \eta \, \Delta t \, \partial^B \phi)\right] \,.
\end{equation}
 which $\tilde{\zeta}$ is the new normalization and $\tilde{P}(x^\prime|x)$ is explicitly covariant.

\section{FOKKER-PLANCK EQUATION}
The integral equation of motion eq. (\ref{rho1-curved}) can be rewritten using the standard method \cite{caticha2012} into a differential equation for the scalar $\rho$ defined as $\mathcal{P} (x,t) = \sqrt{M} \, \rho (x,t)$. Notice that $\rho$ is a scalar field but $\mathcal{P}$ is a tensor density of weight 1.
Then the result of building up a finite change from initial time $t_{0}$ up to final time $t$ by iterating many small changes given by the transition probability $P(x^{\prime}|x)$ is that the probability $\rho$ evolves according a Fokker-Planck (FP) equation
\begin{equation}
\frac{\partial \rho }{\partial t}=-\frac{1}{\sqrt{M}}\partial _{A}\left( 
\sqrt{M}\,\tilde{b}^{A}\rho \right) +\frac{\eta}{2 } \Delta _{M}\, \rho\,,  \label{FP1-curved}
\end{equation}%
 where $\Delta _{M}$ is the Laplace-Beltrami operator
\begin{equation}
\Delta _{M}=\frac{1}{\sqrt{M}}\partial _{A}\left( \sqrt{M}M^{AB}\partial
_{B}\right) \, .
\end{equation}
The FP equation can be further written as a continuity equation
\begin{equation}
\frac{\partial \rho }{\partial t}=-\frac{1}{\sqrt{M}}\partial _{A}\left( 
\sqrt{M} \, v^{A}\rho \right)~,  \label{FP2-curved}
\end{equation}
where $v^A$ is the current velocity
\begin{equation}
v^{A}=\tilde{b}^{A}+u^{A}\,,
\end{equation}
where $u^A$  is the osmotic velocity  
\begin{equation}
u^{A}=- \eta M^{AB}\partial_B \log\rho^{1/2} \,. \label{osmo-curved}
\end{equation}
The continuity equation can also be written as
\begin{equation}
\frac{\partial \rho }{\partial t}=-\frac{1}{\sqrt{M}}\partial _{A}\left( 
\sqrt{M}\rho M^{AB}\partial_{B}\Phi \right)~,  \label{FP3-curved}
\end{equation}
where
\begin{equation}
v^{A}= M^{AB}\partial_{B}\Phi\,\ \ \textrm{with}\quad \Phi(x,t)= \eta \, \phi(x,t)- \eta \, \log\rho^{1/2}(x,t)\,. \label{current-curved}
\end{equation}
Note that since $\rho$ and $\phi$ are scalars (not densities) then $v^A$ and $u^A$ are vectors, and $\Phi$ is a scalar. And finally, for later convenience we write the FP equation as follows
\begin{equation}
\frac{\partial\rho}{\partial t}=\frac{\delta H}{\delta \Phi}\,,\label{H-Phi}
\end{equation}
where $H=H(\Phi,\rho)$. It can easily be checked that the appropriate functional H is 
\begin{equation}
H[\rho,\Phi]=\int\! M^{1/2}(x)d^n x \, \frac{1}{2} \rho M^{AB}\partial_A\Phi\partial_B\Phi+F[\rho]\,,\label{app-H}
\end{equation}
 where $F(\rho)$ is an integration constant to be determined below.

\section{THE SCHR\"{O}DINGER EQUATION IN RIEMANNIAN MANIFOLDS}

The wave function in the Schr\"odinger equation contains two dynamical variable, the probability $\rho$ and phase $\Phi$. But so far we only have one dynamical variable $\rho$ which evolves according to the Fokker-Planck equation (\ref{FP3-curved}). The FP equation tells how to update $\rho$ for an externally given drift potential $\phi$. This is a standard diffusion. In order to promote $\Phi$ to a fully dynamical variable we need to allow the evolving $\rho$ to react back and induce a change in the the potential $\phi$. The precise way in which this reaction is to happen is specified by requiring that there be a conserved ``energy" functional $H=H(\Phi,\rho)$ --- a change in $\rho$ must be compensated by a corresponding change in $\Phi$. 

We require that the energy functional eq. (\ref{app-H}) to be conserved \cite{CBR2014}, 
\begin{equation}
\frac{dH}{dt}=\int\! d^nx\left(\frac{\delta H}{\delta\Phi}\partial_t\Phi+\frac{\delta H}{\delta\rho}\partial_t\rho\right)=0\,,\label{Constant-H1}
\end{equation}
where $\partial_t=\partial/\partial t$. Using eq.~(\ref{H-Phi}) in eq.~(\ref{Constant-H1}) we obtain
\begin{equation}
\frac{dH}{dt}=\int\! d^nx\left(\partial_t\Phi+\frac{\delta H}{\delta\rho}\right)\partial_t\rho=0\,.\label{Constant-H2}
\end{equation}
This must hold for arbitrary initial conditions, or arbitrary $\partial_t\rho$. Therefore, this yields
\begin{equation}
\partial_t\Phi=-\frac{\delta H}{\delta\rho}\,,\label{H-rho}
\end{equation}
for all values of $t$.

Substituting eq. (\ref{app-H}) in eq. (\ref{H-rho}) we get
\begin{equation}
\partial_t \Phi = - \frac{1}{2} M^{AB} \partial_A \Phi \partial_B \Phi - \frac{\delta F}{\delta \rho} \,, \label{HJE}
\end{equation}
which is a generalized Hamilton-Jacobi equation. 

Equations (\ref{FP3-curved}) and  (\ref{HJE}) can be combined into a single equation by introducing a complex function $ \Psi_k$ 
\begin{equation}
\Psi_k = \rho^{1/2} \exp(ik \Phi/\eta) \,,
\end{equation}
where $k$ is a positive constant introduced for later convenience. This yields 
\begin{equation}
\frac{i\eta}{k} \partial_t {\Psi}_k = - \frac{\eta^2}{2 k^2} \Delta_M \Psi_k + \frac{\eta^2}{2 k^2} \frac{\Delta_M | \Psi_k |}{| \Psi_k | } \Psi_k + \frac{\delta {F}}{\delta \rho} \Psi_k \, , \label{nonlinear}
\end{equation}
which is a non-linear Schr\"odinger equation. This can be put into standard Schr\"odinger form by choosing the functional $F(\rho)$ appropriately. Arguments from information geometry suggest that the appropriate choice is \cite{CBR2014}
\begin{equation}
F[\rho] = {\xi} M^{AB} I_{AB} + \int  d^n x \, \rho \, (V + V_c) \label{Feq}\, ,
\end{equation}
where $\xi$ is a coupling constant and $I_{AB}$ is Fisher information metric
\begin{equation}
I_{AB} = \int  d^n x \frac{1}{\rho (x)} \frac{\partial \rho (x)}{\partial x^A} \frac{\partial \rho (x)}{\partial x^B}\, .
\end{equation}
The other terms in eq. (\ref{Feq}) involve the external potential $V$ and a possible curvature potential $V_c$ that vanishes for flat spaces. As we can see the curvature potential only enters through a choice of the scalar $F$ which means that the choice of $V_c$ is arbitrary within the framework of entropic dynamics.

Therefore eq. (\ref{nonlinear}) becomes
\begin{equation}
\frac{i\eta}{k} \partial_t {\Psi}_k = - \frac{\eta^2}{2 k^2} \Delta_M \Psi_k + (\frac{\eta^2}{2 k^2} - 4\xi ) \frac{\Delta_M | \Psi_k |}{| \Psi_k | } \Psi_k + (V + V_c) \Psi_k \, .
\end{equation}
Here we take advantage of the arbitrary constant $k$ and choose it so that $ \xi = \eta^2/8k^2 $. Setting  $\eta /k = \hbar $ we obtain
\begin{equation}
i\hbar \partial_t {\Psi} = - \frac{1}{2} {\hbar}^2 \Delta_M \Psi + (V+V_c) \, \Psi \, ,
\end{equation}
which is the Schr\"odinger equation for $N$ particles on the curved configuration space. In flat space we have $\Delta_M \to \sum_{i=1}^{N} \nabla^2_i /m_i$, the curvature potential vanishes and we recover the usual Shr\"odinger equation,
\begin{equation}
i\hbar\partial_t {\Psi} = - \sum_{i} \frac{\hbar^2}{2 m_i} \nabla^2_i \Psi + V \Psi \, .
\end{equation}

\section{SUMMARY}
Entropic Dynamics views quantum theory as an application of entropic methods of inference. In entropic dynamics no underlying action principle is assumed, on the contrary, an action principle and the corresponding Hamiltonian dynamics are derived as a non-dissipative diffusion.

Our goal has been to extend entropic dynamics to curved spaces which is an important preliminary step toward an entropic dynamics of gravity. We have derived the modified Schr\"{o}dinger equation on a Riemannian manifold in the framework of entropic dynamics. The modified equation replaces the Laplacian by the Laplace-Beltrami operator.

\section*{ACKNOWLEDGMENTS}
We would like to thank C. Cafaro, D. Bartolomeo, S. DiFranzo, S. Ipek, K. Vanslette, A. Yousefi, A. Fernandes and P. Pessoa for many insightful discussions.
\providecommand{\href}[2]{#2}

\end{document}